\documentclass[%
 reprint,
superscriptaddress,
preprintnumbers,
 amsmath,amssymb,
 prl,
]{revtex4-2}

\usepackage{graphicx}
\usepackage{dcolumn}
\usepackage{bm}

\usepackage{float}
\usepackage{hyperref}
\usepackage{amsmath}
\usepackage{amssymb}
\usepackage{microtype}
\usepackage{cancel}
\usepackage[normalem]{ulem}
\usepackage{amsbsy}
\usepackage{color}
\usepackage{breakurl}
\usepackage{mathtools}
\usepackage{lmodern}
\usepackage{dsfont}
\usepackage{bbold}
\usepackage{empheq}
\usepackage{enumitem}
\usepackage{bbm}
\usepackage{comment}
\usepackage{accents}
\usepackage[dvipsnames, svgnames]{xcolor}
\usepackage{empheq}
\usepackage[capitalise]{cleveref}

\DeclareMathAlphabet{\mathpzc}{OT1}{pzc}{m}{it}

\newcommand{\para}[1]{\par\vspace{2mm}\noindent\textbf{#1}\,---\,}

\makeatletter
\DeclareRobustCommand{\rcite}[1]{%
  \rcite@aux#1,\@nil{#1}%
}
\def\rcite@aux#1,#2\@nil#3{%
  \if\relax#2\relax
    Ref.~\cite{#3}%
  \else
    Refs.~\cite{#3}%
  \fi
}
\makeatother

\hypersetup{
    colorlinks = true,
    citecolor = {red},
    linkcolor = {blue},
    urlcolor = {blue},
}

\begin{document}

\title{Quantum Signatures of Cosmic Topology: \\ How Casimir Backreaction Transmits Isotropy Violation}

\author{Anna Negro}
\email{anna.negro@case.edu}
\affiliation{CERCA/ISO, Department of Physics, Case Western Reserve University, Cleveland, Ohio 44106, USA}

\author{Kurt Hinterbichler}
\affiliation{CERCA/ISO, Department of Physics, Case Western Reserve University, Cleveland, Ohio 44106, USA}

\author{Glenn D. Starkman}
\affiliation{CERCA/ISO, Department of Physics, Case Western Reserve University, Cleveland, Ohio 44106, USA}

\author{Yashar~Akrami}
\affiliation{Instituto de F\'isica Te\'orica (IFT) UAM-CSIC, C/ Nicol\'as Cabrera 13-15, Campus de Cantoblanco UAM, 28049 Madrid, Spain}
\affiliation{CERCA/ISO, Department of Physics, Case Western Reserve University, Cleveland, Ohio 44106, USA}
\affiliation{Astrophysics Group \& Imperial Centre for Inference and Cosmology, Department of Physics, Imperial College London, Blackett Laboratory, Prince Consort Road, London SW7 2AZ, United Kingdom}

\author{Stefano~Anselmi}
\affiliation{INFN, Sezione di Padova, via Marzolo 8, I-35131 Padova, Italy}
\affiliation{Dipartimento di Fisica e Astronomia ``G. Galilei'', Universit\`a degli Studi di Padova, via Marzolo 8, I-35131 Padova, Italy}
\affiliation{Laboratoire Univers et Th\'eories, Observatoire de Paris, Universit\'e PSL, Universit\'e Paris Cit\'e, CNRS, F-92190 Meudon, France}

\author{Javier~Carr\'on~Duque}
\affiliation{Instituto de F\'isica Te\'orica (IFT) UAM-CSIC, C/ Nicol\'as Cabrera 13-15, Campus de Cantoblanco UAM, 28049 Madrid, Spain}

\author{Mikel~Martin~Barandiaran}
\affiliation{Instituto de F\'isica Te\'orica (IFT) UAM-CSIC, C/ Nicol\'as Cabrera 13-15, Campus de Cantoblanco UAM, 28049 Madrid, Spain}
\affiliation{Departamento de F\'isica Te\'orica, Universidad Aut\'onoma de Madrid, 28049 Madrid, Spain}

\author{Thiago~S.~Pereira}
\affiliation{Departamento de F\'{i}sica, Universidade Estadual de Londrina, Rod. Celso Garcia Cid, Km 380, 86057-970, Londrina, Paran\'{a}, Brazil}

\author{George Alestas}
\affiliation{Instituto de F\'isica Te\'orica (IFT) UAM-CSIC, C/ Nicol\'as Cabrera 13-15, Campus de Cantoblanco UAM, 28049 Madrid, Spain}

\author{Craig~J.~Copi}
\affiliation{CERCA/ISO, Department of Physics, Case Western Reserve University, Cleveland, Ohio 44106, USA}

\author{Fernando~Cornet-Gomez}
\affiliation{University of Cordoba, Dept. of Physics, Campus de Rabanales, Alfonso XIII, 13, E-14071 Cordoba, Spain}

\author{Linn~Htat~Lu}
\affiliation{Astrophysics Group \& Imperial Centre for Inference and Cosmology, Department of Physics, Imperial College London, Blackett Laboratory, Prince Consort Road, London SW7 2AZ, United Kingdom}

\author{Andrew~H.~Jaffe}
\affiliation{Astrophysics Group \& Imperial Centre for Inference and Cosmology, Department of Physics, Imperial College London, Blackett Laboratory, Prince Consort Road, London SW7 2AZ, United Kingdom}

\author{Arthur~Kosowsky}
\affiliation{Department of Physics and Astronomy, University of Pittsburgh, Pittsburgh, Pennsylvania 15260, USA}

\author{Deyan~P.~Mihaylov}
\affiliation{CERCA/ISO, Department of Physics, Case Western Reserve University, Cleveland, Ohio 44106, USA}
\affiliation{Department of Astronomy, Faculty of Physics, Sofia University ``St.~Kliment~Ohridski'', 5 James Bourchier Blvd, 1164 Sofia, Bulgaria}

\author{Joline~Noltmann}
\affiliation{Instituto de F\'isica Te\'orica (IFT) UAM-CSIC, C/ Nicol\'as Cabrera 13-15, Campus de Cantoblanco UAM, 28049 Madrid, Spain}
\affiliation{Institute for Theoretical Particle Physics and Cosmology, RWTH Aachen University, Templergraben 55, 52062 Aachen, Germany}

\author{Jos\'e~Javier~Ortega~G\'omez}
\affiliation{Instituto de F\'isica Te\'orica (IFT) UAM-CSIC, C/ Nicol\'as Cabrera 13-15, Campus de Cantoblanco UAM, 28049 Madrid, Spain}
\affiliation{Departamento de F\'isica Te\'orica, Universidad Aut\'onoma de Madrid, 28049 Madrid, Spain}

\author{Catherine~Petretti}
\affiliation{Harvard-Smithsonian Center for Astrophysics (CfA), 60 Garden St., Cambridge, MA 02138, USA}

\author{Amirhossein~Samandar}
\affiliation{CERCA/ISO, Department of Physics, Case Western Reserve University, Cleveland, Ohio 44106, USA}

\author{Andrius~Tamosiunas}
\affiliation{Institute of Theoretical Astrophysics, P.O. Box 1029 Blindern, N-0315 Oslo, Norway}

\collaboration{COMPACT Collaboration}

\date{\today}

\begin{abstract}
A finite, scheme-independent Casimir contribution to the stress-energy tensor arises naturally for quantum fields in universes with non-trivial spatial topology. 
We compute this Casimir stress-energy tensor contribution for a conformally coupled scalar field and for a minimally coupled scalar field. 
We show that, for the conformally coupled case, the backreaction of this contribution to the Einstein equations during an expanding de Sitter phase drives anisotropic expansion even when the Universe begins in a locally homogeneous and isotropic state.
We conclude that quantum imprints of the underlying non-trivial topology inevitably give rise to local departures from homogeneity and isotropy. 
\end{abstract}

\preprint{IFT-UAM/CSIC-26-25}

\maketitle

\para{Introduction.}
Whether our Universe has non-trivial spatial topology has been a long-standing question \cite{1900VAGes..35..337S,Sommerville} (see Ref. \cite{COMPACT:2026nez} for a review). Different manifolds can share the same local geometry yet have different topologies. Therefore, assuming that the Universe is well described by a spatially flat, spherical, or hyperbolic Friedmann-Lema\^{i}tre-Robertson-Walker (FLRW) solution does not uniquely determine the cosmic topology and, consequently, the shape of our Universe. At present, there is no settled first-principles prediction for either cosmic topology or its characteristic scales. At the same time, studies of quantum creation suggest that non-trivial topology is a viable possibility \cite{Linde:2004nz,Hassfeld:2024hnx,Carlip:2022pyh,Hawking:1978pog,Hassfeld:2024ake,Godet:2026gqw}. In this work we therefore study non-trivial cosmic topology as a theoretically motivated possibility, while treating the specific topology and its associated length scale as phenomenological inputs to be assessed through their observational signatures.

\color{black}{The} search for cosmic topology has so far focused on late-time phenomenology---namely, the observability of correlations in the cosmic microwave background (CMB) or in the large-scale structure of the Universe.
In this work, we show that quantum vacuum effects induced by non-trivial topology can drive a period of anisotropic expansion at the onset of inflation, potentially producing observable effects on inflationary perturbations and providing a new avenue for probing cosmic topology.

In non-trivial topologies points of the simply connected covering space are identified under a discrete group of isometries. If the identification scale is smaller than the diameter of the last-scattering surface (LSS), the observer’s past light cone intersects itself, generating pairs of matched circles in the CMB \cite{Cornish:1997ab}. To date, searches for such ``circles in the sky" have not found statistically significant evidence \cite{deOliveira-Costa:2003utu,Cornish:2003db,ShapiroKey:2006hm,Mota:2010jb,Bielewicz:2012jnb,Bielewicz_2012,PhysRevD.86.083526,Aurich:2013fwa,Planck:2013okc,Planck:2015gmu,Planck:2018vyg}, placing an upper bound on the shortest closed path. Crucially, such constraints do not rule out non-trivial topology altogether and cosmic topology
remains an active area of research \cite{starobinsky1993newrestrictionsspatialtopology,PhysRevLett.71.20,Roukema:2006yd,Vigneron_2023,Fabre:2013wia,Aurich:2021ofm,Phillips:2004nc,Niarchou:2007nn,COMPACT:2022gbl}. For spatially flat FLRW solutions, promising signatures of cosmic topology follow from the classification of the $18$ Euclidean topologies ($17$ non-trivial topologies and the simply connected covering space), each characterized by a fundamental domain together with specific topological boundary conditions \cite{COMPACT:2023rkp,COMPACT:2025adc,Luminet:1999qh,Lachieze-Rey:1995qrb,PhysRevD.69.103518}.

\begin{figure}[b]
\includegraphics[width=0.45\textwidth]{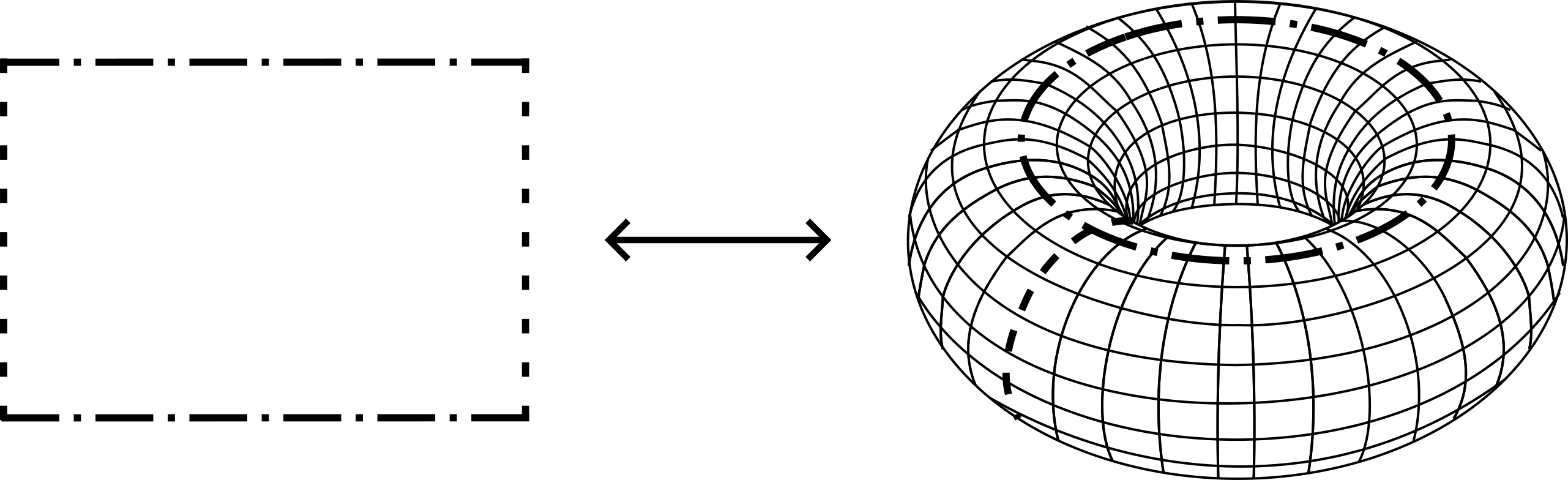}
\caption{\footnotesize{A two-dimensional torus is obtained by boundary conditions identifying the opposite sides of a rectangular fundamental domain. Similarly, in three-dimensional Euclidean space, each of the 17 non-trivial topologies can be represented by topological identifications imposed on a fundamental domain \cite{Riazuelo:2003ud}.}}
\label{fig:Topo1}
\end{figure}

\begin{figure}[hpt!]
\includegraphics[width=0.3\textwidth]{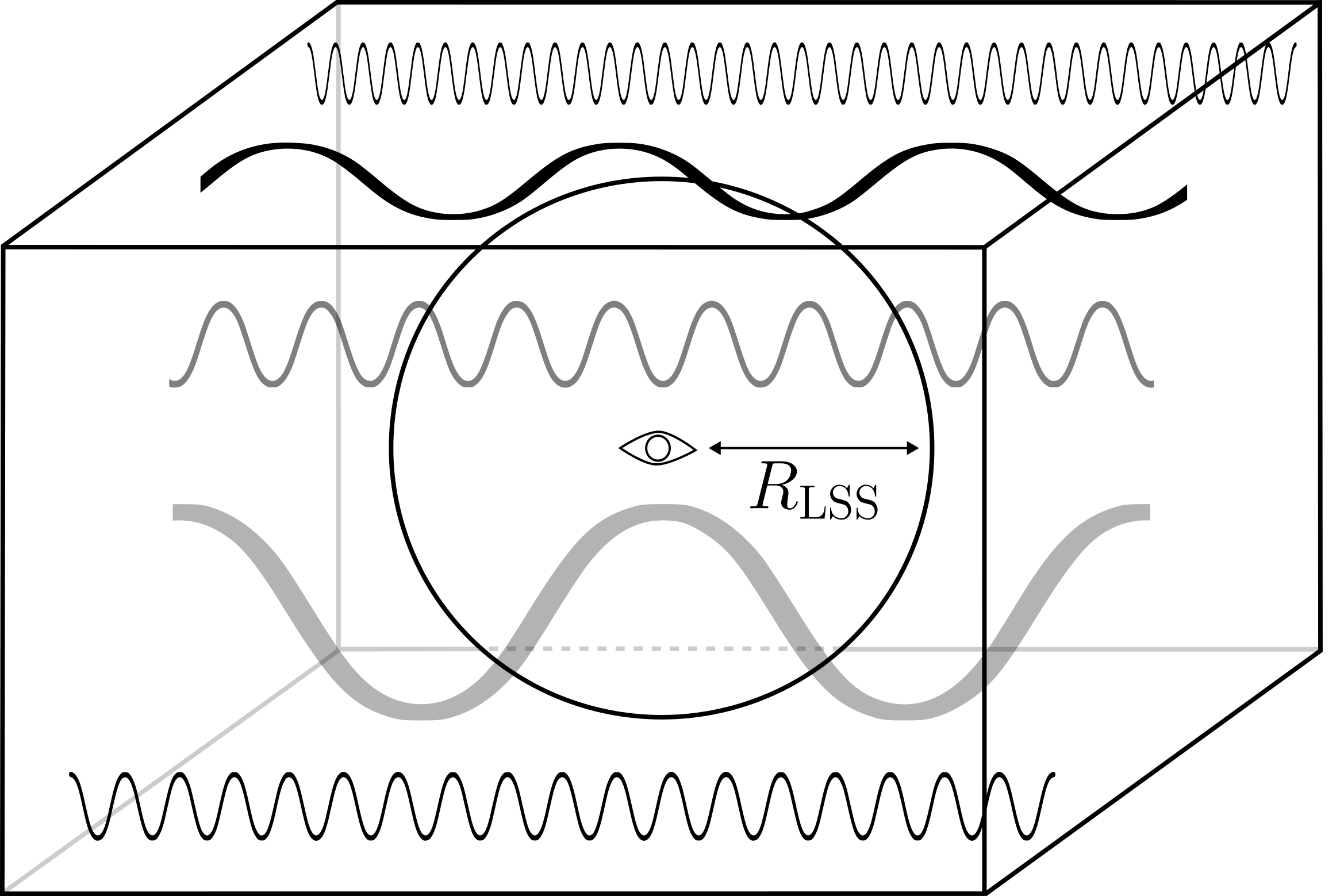}
\caption{\footnotesize{A simply connected space allows fluctuations of arbitrary wavelength on the last scattering surface, the furthest optically observable region with radius $R_{\mathrm{LSS}}$. By contrast, non-trivial topology, such as a three-dimensional cubic torus with elongated fundamental domain, restricts the admissible modes to a discrete set fixed by its boundary conditions. For the torus, only modes satisfying periodic boundary conditions contribute to the statistics of CMB fluctuations, even when the characteristic size of the topology is bigger than $R_{\mathrm{LSS}}$.}}
\label{fig:Topo3}
\end{figure}

Representing a non-trivial topology using a fundamental domain and its associated boundary conditions, as illustrated in \cref{fig:Topo1}, makes explicit how global topology introduces compact dimensions and constrains physical fields beyond matched-circle pairs. As shown in \cref{fig:Topo3}, the allowed wavelengths are constrained both by the finite extent of the fundamental domain and by the topological boundary conditions, selecting only modes that are invariant under the action of the discrete isometry group defining the underlying topology. Moreover, while the simply connected covering space admits translational and rotational isometries, the identifications defining the $17$ non-trivial topologies inevitably break global isotropy and, in some cases, global homogeneity. Building on this framework, the COMPACT collaboration \cite{COMPACT:2022gbl} has been investigating the constraining power of CMB perturbations, finding that if the shortest closed path around the Universe through the observer is shorter than \(\sim 1.05-1.1\) times the diameter of the LSS, the CMB contains sufficient distinguishing information in temperature and polarization to identify whether our Universe is topologically trivial or non-trivial \cite{Samandar:2025kuf,COMPACT:2024qni,COMPACT:2024cud,COMPACT:2022nsu}.

Existing efforts, however, focus on identifying observational imprints of cosmic topology when the dynamics are well described classically. By contrast, considerably less attention has been paid to the role of non-trivial spatial topology during the early Universe, when quantum effects are expected to play a central role. The limited work in this direction suggests that topology can have substantial effects in the quantum regime. Namely, effects of non-trivial topology in the early Universe have been shown to generate homogeneity \cite{Cornish:1996st,Barrow:1997sb,Barrow:1999sm}, modify quantum-particle spectra and particle production \cite{Paraskevas:2025cwu,Fornal:2012kx}, and improve the prospects for detecting cosmic topology \cite{Noltmann:2026ise}.

In the early Universe, gravity is typically treated semiclassically, and a direct consequence of non-trivial spatial topology in such a framework is the emergence of a Casimir stress-energy tensor. 
Originally identified in the context of conducting plates \cite{Casimir:1948dh}, the Casimir effect is more generally a pure vacuum contribution arising when boundary conditions modify the energy eigenspectrum. In particular, it is a natural consequence of the mode selection imposed by non-trivial topology, as illustrated in \cref{fig:Topo3}. 
Although this effect is well understood \cite{Bire82,DeWitt:1979dd,DeWitt:1975ys} and has been explored in a variety of settings---including semiclassical bouncing cosmologies \cite{Herdeiro:2005zj,Godlowski:2007gx,Alexandre:2019ygz} and higher-dimensional theories, where Casimir energies influence moduli or brane dynamics \cite{Fabinger:2000jd,Nasri:2002rx,Greene:2007xu,Jacobs:2012id,Jacobs:2012ph,DeLuca:2021pej}---its role as a direct imprint of non-trivial spatial topology in the early Universe has remained unexplored.

We proceed below to begin to remedy this important gap in our understanding of the phenomenological implications of cosmic topology. We demonstrate that quantum vacuum effects in topologically non-trivial manifolds generate scheme-independent Casimir contributions that locally encode the global symmetry breaking imposed by the underlying topology. We show how non-trivial topology affects the vacuum-state selection, resulting in isotropic zero-mode contributions that can modify the effective equation of state of the total matter content. Additionally, in the three-torus case, we find that the Casimir contribution can source anisotropic expansion in the early Universe, with effects that could be observable to a late-time observer.

As a specific illustration of these effects, we derive the Casimir stress-energy tensor for both a massless and a conformally coupled scalar field on de Sitter space, in which the spatial slices of the inflationary patch have a toroidal topological identification. These serve as representative examples of very light degrees of freedom, such as the inflaton and spectator fields with Hubble scale masses, that may be present during inflation. We then treat the backreaction perturbatively by including the Casimir stress-energy tensor of the conformally coupled field in the Einstein equations and solving for the first-order corrections to the metric, obtaining an anisotropic correction that leads to direction-dependent expansion rates. Our results therefore identify Casimir backreaction as a quantum signature of non-trivial cosmic topology that transmits local anisotropy in the early Universe.

\para{Casimir stress-energy tensor.}
We first derive the Casimir stress-energy tensor for a conformally coupled scalar field on a de Sitter expanding background. This choice permits an explicit analytic treatment while cleanly isolating the backreaction effects associated with spatial topology. Since the Casimir contribution originates from the restriction of vacuum modes by compact spatial dimensions, the mechanism is general and not restricted to the specific field content or coupling. We therefore briefly comment on the corresponding minimally coupled result, relevant for cosmological scalars such as the inflaton, leaving its detailed analysis to future work.

We start from the action for a conformally coupled scalar field $\phi$,
\begin{equation}\label{eq:action}
    S=\frac{1}{2}\int \mathrm{d}^4x\sqrt{-g}
    \left[-\nabla_\mu\phi\nabla^\mu\phi-\frac{1}{6}\phi^2R\right]\,,
\end{equation}
where $R$ is the Ricci scalar. We take the background to be de Sitter, written in conformal time as
\begin{equation}\label{eq:bkmetric}
    g_{\mu \nu}=\operatorname{diag}\left(-a(\eta)^2,a(\eta)^2,a(\eta)^2,a(\eta)^2\right)\,,
\end{equation}
for which the scale factor is $a(\eta)=-\frac{1}{H\eta}$, with $\eta\in(-\infty,0)$ and $H$ the Hubble expansion rate. We take the spatial topology to be a three-dimensional torus with a general fundamental domain.\color{black}{}
Consequently, the scalar field decomposition in the basis of the  eigenmodes of the Helmholtz equation is characterized by a discrete set of momenta $\vec{k}_{\vec{n}}$ that are fixed by the associated topological boundary conditions
\begin{equation}\label{eq:kn}
    \vec{k}_{\vec{n}} \cdot \vec{T}_j = 2 \pi n_j\,, \quad \forall n_j \in \mathbb{Z}\,,\quad j=1,2,3 \,.
\end{equation}
Here $\{\vec{T}_1,\vec{T}_2,\vec{T}_3\}$ are the translation vectors that generate the lattice defining the fundamental domain. 
For example, for a rectangular lattice, $\vec{T}_j=L\beta_j\hat{e}_j$, where $L$ denotes the characteristic size of the fundamental domain and $\beta_j$ are dimensionless parameters specifying its shape, giving 
\begin{equation}\label{eq:knrect}
    \vec{k}_{\vec{n}}= \frac{2\pi}{L} \left(\frac{n_1}{\beta_1}, \frac{n_2}{\beta_2},\frac{n_3}{\beta_3} \right) \,.
\end{equation} 
The field can therefore be written as
\begin{equation}\label{eq:field}
    \phi (\eta,\vec{x}) = \frac{1}{\mathrm{V}} \:  \sum_{\vec{n}} \:\left[ a_{\vec{n}}  e^{i \vec{k}_{\vec{n}}\cdot \vec{x}} \varphi_{\vec{n}}(\eta)+a^\dagger_{\vec{n}} e^{-i \vec{k}_{\vec{n}}\cdot \vec{x}} \varphi_{\vec{n}}^*(\eta) \right] \,,
\end{equation}
where $\mathrm{V}=\vert\vec{T}_1\cdot(\vec{T}_2\times\vec{T}_3)\vert$ is the volume of the fundamental domain, and $a_{\vec{n}}$, $a^\dagger_{\vec{n}}$ are the creation and annihilation operators, respectively, normalized as $[a_{\vec{n}},a^\dagger_{\vec{m}}]=\mathrm{V}\delta_{\vec{n},\vec{m}}$. 
The mode functions satisfy the Klein-Gordon equation and  take the form 
\begin{subequations}
\label{eq:modes}
\begin{align}
\label{eq:modesn}
&  \mkern-8mu\varphi_{\vec{n}\neq \vec{0}} =H \vert \eta \vert^\frac{3}{2}  \left[ b_{\vec{n}}  H^{(1)}_{\frac{1}{2}} (-|\vec{k}_{\vec{n}}| \eta)+c_{\vec{n}}  H^{(2)}_{\frac{1}{2}} (-|\vec{k}_{\vec{n}}| \eta)\right]\,, \\
\label{eq:modes0}
&  \mkern-8mu\varphi_{\vec{0}} = \sqrt{H} b_{\vec{0}} \eta  + \sqrt{H^3} c_{\vec{0}} \eta^2 \,,
\end{align}
\end{subequations}
where $H^{(1)}_{\frac{1}{2}} $ and $H^{(2)}_{\frac{1}{2}}$ are Hankel functions of the first and second kind, and where $b_{\vec{n}}$, $c_{\vec{n}}$, $b_{\vec{0}}$, and $c_{\vec{0}}$ are integration constants.

The integration constants of the mode functions in \cref{eq:modes} are fixed by a choice of vacuum state. For the oscillatory modes we choose the Bunch-Davies (BD) prescription \cite{Bunch:1978yw,Bunch:1977sq}, which amounts to requiring positive-frequency plane-wave behavior at early times. For $\vec{n}\neq\vec{0}$, this uniquely fixes $b_{\vec{n}}=-i\frac{\sqrt{\pi}}{2}$ and $c_{\vec{n}}=0$, ensuring that the quantum state reduces to the Minkowski vacuum when the physical wavelength is much smaller than the Hubble radius. This prescription also fixes the ultraviolet behavior of the state and ensures the Hadamard form, both in the compact case and in the simply connected covering space. The zero mode in \cref{eq:modes0}, however, is not fixed by this requirement and represents a residual freedom left unfixed by the Hadamard condition. In the simply connected case, $b_{\vec{0}}$ and $c_{\vec{0}}$ are fixed by requiring the two-point function to be de Sitter invariant, thereby selecting the BD vacuum. In the case under analysis, however, the topology-dependent image contribution already breaks the de Sitter invariance of the two-point function. The same symmetry requirement cannot be used to set $b_{\vec{0}}=c_{\vec{0}}=0$ and discard the zero-mode contribution. We therefore leave $b_{\vec{0}}$ and $c_{\vec{0}}$ unspecified; their physical significance will be commented later.

We now have all the necessary ingredients to compute the stress-energy tensor. 
We do this by varying the action in \cref{eq:action} with respect to the metric and taking the expectation value, denoted by $\langle\cdot\rangle$, we find $\langle T_{i0}\rangle=0=\langle T_{0j}\rangle$, and
\begin{subequations}
\label{eq:T1}
\begin{align}\label{eq:rho1}
&\rho =-\langle T_{0}{}^{0}\rangle = \frac{H^4 \eta^4}{\mathrm{V}} \sum_{\vec{n} \in \mathbb{Z}^3 \setminus \{\vec{0}\}}  \frac{|\vec{k}_{\vec{n}}|}{2}\ + \frac{H^5 \eta^4}{2 \mathrm{V}}  |c_{\vec{0}}|^2\,, \\
\label{eq:P1}
&P_{i}{}^{j} =\langle T_{i}{}^{j}\rangle =\frac{H^4 \eta^4}{\mathrm{V}} \mkern-8mu \sum_{\vec{n} \in \mathbb{Z}^3 \setminus \{\vec{0}\}} \mkern-8mu \frac{(\vec{k}_{\vec{n}})_i (\vec{k}_{\vec{n}})^j}{2|\vec{k}_{\vec{n}}|}+ \frac{H^5  \eta^4}{6 \mathrm{V}} |c_{\vec{0}}|^2  \delta_{i}{}^{j} \,. 
\end{align}
\end{subequations}
We observe that the stress-energy tensor is independent of $b_{\vec{0}}$, while its dependence on $c_{\vec{0}}$ contributes an isotropic term. 
To isolate the contribution due to non-trivial topology, we first include the zero mode in the sum in \cref{eq:T1} and then apply the Poisson summation formula \cite{cdadf853-2bd9-3849-9928-547f47a288f0}, which relates the sum over the reciprocal lattice $\{\vec{k}_{\vec{n}}\}$ defined in \cref{eq:kn} to a sum over the direct lattice of images,
\begin{equation}\label{eq:lr}
    \vec{l}_{\vec{r}} = \sum_{j=1}^3 r_j \vec{T}_{j} \,, \quad 
    r_j \in \mathbb{Z} \,.
\end{equation}
This representation naturally separates the vacuum expectation value into the simply connected covering-space contribution, corresponding to the $\vec{r}=0$ term in position space, and a topology-dependent correction. 
Subtracting the former ($\rho^{\mathrm{cs}}$, $P_{ij}^{\mathrm{cs}}$), which is ultraviolet divergent due to the coincidence limit in the definition of the stress-energy tensor \cite{PhysRevD.17.946,1978RSPSA.360..117B,Dowker:1975tf}, we obtain the finite, topology-induced contributions
\begin{subequations}
    \label{eq:T2}
    \begin{align}\label{eq:rho2}
    &\rho^{\mathrm{tp}} \equiv \rho -\rho^{\mathrm{cs}}= - \frac{H^4 \eta^4}{2\pi^2}\mkern-6mu \sum_{\vec{r} \in \mathbb{Z}^3 \setminus \{\vec{0}\}}  \mkern-6mu \frac{1}{ |\vec{l}_{\vec{r}}|^4} + \frac{H^5 \eta^4}{2 \mathrm{V}}  |c_{\vec{0}}|^2 \,,
    \\
    \label{eq:P2}
    &P^{\mathrm{tp}}{}_{i}{}^{j}  \equiv P_{i}{}^{j}-P^{\mathrm{cs}}{}_{i}{}^{j}\\ \nonumber
    & \mkern40mu = \mkern-2mu \frac{H^4 \eta^4}{2 \pi^2 } \mkern-14mu \sum_{\vec{r}  \in \mathbb{Z}^3 \setminus \{\vec{0}\}}  \mkern-14mu \frac{|\vec{l}_{\vec{r}}|^2 \delta_{i}{}^{j} \! - \! 4 (\vec{l}_{\vec{r}})_i  (\vec{l}_{\vec{r}})^j }{|\vec{l}_{\vec{r}}|^6}  +  \frac{H^5  \eta^4 |c_{\vec{0}}|^2  \delta_{i}{}^{j}}{6 \mathrm{V}} \,.
    \end{align}
\end{subequations}

Following the same general procedure for a minimally coupled scalar field, one finds $\langle T_{i0}\rangle=0=\langle T_{0j}\rangle$,
\begin{subequations}
\label{eq:Tminim1}
\begin{eqnarray}
\label{eq:rhomin1}
&& \mkern-12mu \rho =  \frac{H^4 \eta^2}{ \mathrm{V}} \mkern-12mu \sum_{\vec{n} \in \mathbb{Z}^3 \setminus \{\vec{0}\} } \mkern-4mu \left[  \frac{\eta^2 |\vec{k}_{\vec{n}}|}{2} + \frac{1}{4 |\vec{k}_{\vec{n}}|} \right] + \frac{ H^7 \eta^6}{2 \mathrm{V}}  |c_{\vec{0}}|^2 \,,
\\ \nonumber
&&  \mkern-12mu P_{i}{}^{j}  = \mkern-4mu \frac{H^4 \eta^2}{  \mathrm{V}} \mkern-20mu \sum_{\vec{n} \in \mathbb{Z}^3 \setminus \{\vec{0}\} } \mkern-6mu \left[ \frac{\eta^2 (\vec{k}_{\vec{n}})_i (\vec{k}_{\vec{n}})^j}{2 |\vec{k}_{\vec{n}}|}+  \frac{ (\vec{k}_{\vec{n}})_i (\vec{k}_{\vec{n}})^j}{2 |\vec{k}_{\vec{n}}|^3}-  \frac{\delta_{i}{}^{j}}{4 |\vec{k}_{\vec{n}}|} \right]\\
&&\mkern40mu+ \frac{ H^7 \eta^6}{2 \mathrm{V}}  |c_{\vec{0}}|^2 \delta_{i}{}^{j}\,.
\end{eqnarray}
\end{subequations}
We next isolate the topology-dependent contribution by applying the Poisson summation formula. Since this step requires temporarily extending the sums to the full lattice, we use the Ewald summation technique \cite{NIJBOER1957309,ValeixoBento:2025yhz} to track the artificially introduced divergent zero-mode contribution and recover the original infrared-finite result in position space. \color{black}{In} particular, for the sum $\sum_{\vec{r}\in\mathbb{Z}^3\setminus\{\vec{0}\} }|\vec{l}_{\vec{r}}|^{-2}$ we obtain
\begin{eqnarray}\label{eq:Ewald}\nonumber
     \mkern-12mu \widetilde{\sum_{\vec{r} \in \mathbb{Z}^3 \setminus \{\vec{0}\} }}   \frac{1}{|\vec{l}_{\vec{r}}|^2}   \equiv&&  \mkern-20mu\sum_{\vec{r} \in \mathbb{Z}^3 \setminus \{\vec{0}\}} \frac{e^{-\lambda |\vec{l}_{\vec{r}}|^2}}{|\vec{l}_{\vec{r}}|^2}-\frac{2 \pi^{3 / 2}}{\operatorname{V} } \frac{1}{\sqrt{\lambda}} - \lambda\\
    && \mkern-20mu+\frac{2 \pi^2}{\operatorname{V}} \mkern-8mu \sum_{\vec{n} \in \mathbb{Z}^3 \setminus \{\vec{0}\}} \frac{1}{\left|k_{\vec{n}}\right|} \operatorname{erfc}\left(\frac{\left|k_{\vec{n}}\right|}{2 \sqrt{\lambda}}\right) \, ,
\end{eqnarray}
where the tilde over the summation symbol denotes the finite Ewald-summed lattice contribution defined by the convergent expression on the right-hand side. The topology-dependent part of the Casimir stress-energy tensor for a minimally coupled field therefore becomes 
\begin{subequations}
\label{eq:Tminim2}
\begin{align}
&\rho^{\mathrm{tp}}\mkern-2mu
= \mkern-4mu - \frac{H^4 \eta^4}{2\pi^2}
\mkern-16mu \sum_{\vec{r} \in \mathbb{Z}^3 \setminus \{\vec{0}\}}
\mkern-6mu \frac{1}{ |\vec{l}_{\vec{r}}|^4}
\mkern-2mu + \mkern-2mu \frac{H^4 \eta^2 }{ 8 \pi^2 } \mkern-10mu
\widetilde{\sum_{\vec{r} \in \mathbb{Z}^3 \setminus \{\vec{0}\}}}\mkern-2mu
\frac{1}{|\vec{l}_{\vec{r}}|^2} \mkern-2mu + \mkern-2mu \frac{ H^7 \eta^6}{2 \mathrm{V}}  |c_{\vec{0}}|^2\,,
\label{eq:rhomin2}
\\[-1ex]
&P^{\mathrm{tp}}{}_{i}{}^{j}\mkern-2mu
=\mkern-2mu
\frac{H^4 \eta^4}{2 \pi^2 } \mkern-18mu
\sum_{\vec{r}  \in \mathbb{Z}^3 \setminus \{\vec{0}\}} \mkern-20mu
\frac{
|\vec{l}_{\vec{r}}|^2 \delta_{i}{}^{j}
\mkern-2mu - \mkern-2mu4 (\vec{l}_{\vec{r}})_i  (\vec{l}_{\vec{r}})^j
}
{|\vec{l}_{\vec{r}}|^6}
\mkern-2mu + \mkern-2mu \frac{H^4 \eta^2 }{ 8 \pi^2 }
\mkern-10mu \widetilde{\sum_{\vec{r} \in \mathbb{Z}^3 \setminus \{\vec{0}\} }}
\frac{\delta_{i}{}^{j}}{|\vec{l}_{\vec{r}}|^2}
\nonumber\\
&\mkern40mu
-\frac{H^4 \eta^2 }{ 2 \pi^2 }
\mkern-10mu
\widetilde{\sum_{\vec{r} \in \mathbb{Z}^3 \setminus \{\vec{0}\}}}\mkern-4mu
 \frac{ (\vec{l}_{\vec{r}})_i  (\vec{l}_{\vec{r}})^j }
{|\vec{l}_{\vec{r}}|^4} + \frac{H^7 \eta^6}{2 \mathrm{V}}
|c_{\vec{0}}|^2 \delta_{i}{}^{j} \,.
\label{eq:Pmin2}
\end{align}
\end{subequations}
where the second-last term in \cref{eq:Pmin2} follows from differentiating the Ewald representation of the lattice sum in \cref{eq:Ewald} with respect to the parameters defining the fundamental domain. 

As in the conformally coupled case, the zero-mode contribution to the stress-energy tensor enters through an isotropic term proportional to $c_{\vec{0}}$, whose value is set by the vacuum choice of the zero mode.
We highlight that both the results in \cref{eq:T2} and \cref{eq:Tminim2} show that the topology-dependent contribution is finite and can be extracted by taking the coincidence limit at the outset. 
In contrast, to correctly reproduce the trace anomaly \cite{Duff:1993wm,Capper:1974ic,Brown:1976wc,Brown:1977sj,Wald:1978pj}, $\rho^{\mathrm{cs}}$ and $P_{ij}^{\mathrm{cs}}$ must be regularized prior to taking the coincidence limit. 
This demonstrates that the Casimir contribution is finite and scheme-independent, while counterterm ambiguities affect only the local, covering-space part.

\para{Backreaction.} Focusing on the conformally coupled case, we now use the Casimir contribution derived in \cref{eq:T2} to quantify its effects on inflation. 
We proceed to consider a generic rectangular fundamental domain, $\vec{l}_{\vec{r}}=L(\beta_{1}r_1,\beta_{2}r_2,\beta_{3}r_3)$, and include our result for the Casimir stress-energy tensor,
\begin{widetext}
\begin{equation}\label{eq:Tmunu} 
\langle T^{\mathrm{tp}}{}_{\mu}{}^{\nu}\rangle =\frac{H^4 \eta^4}{2\pi^2  }  \sum_{\vec{r}  \in \mathbb{Z}^3 \setminus \{\vec{0}\}}   \frac{1}{ |\vec{l}_{\vec{r}}|^4} \operatorname{diag}\left(1,1 - 4 \frac{L^2 \beta_1^2 r_1^2}{|\vec{l}_{\vec{r}}|^2} ,1 - 4 \frac{L^2 \beta_2^2r_2^2}{|\vec{l}_{\vec{r}}|^2} , 1 - 4 \frac{L^2 \beta_3^2 r_3^2}{|\vec{l}_{\vec{r}}|^2} \right) + \frac{H^5  \eta^4 |c_{\vec{0}}|^2 }{2 L^3 | \beta_1 \beta_2 \beta_3 |} \operatorname{diag}\left(-1, \frac{1}{3},\frac{1}{3},\frac{1}{3}  \right)\,,
\end{equation}
\end{widetext}
on the right-hand side of the Einstein equations. We note that the off-diagonal components vanish by reflection symmetry, and that the locally anisotropic structure of $\langle T^{\mathrm{tp}}{}_{\mu}{}^{\nu}\rangle$, which is most significant in the early Universe, when the physical topological scale is small, explicitly reflects the globally reduced symmetry of the underlying topology. The term proportional to $c_{\vec{0}}$ instead generates an isotropic radiation-like contribution, which is physical and depends on the vacuum choice of the zero mode. Similarly, if the manifold possesses an  accidental cubic symmetry ($\beta_1=\beta_2=\beta_3$), one finds the standard, isotropic, radiation-like Casimir stress-energy tensor of the cubic torus. These contributions enter in the right-hand side of Einstein’s equations and can affect the effective equation of state of the total stress-energy content. Since our interest is in the anisotropic contribution sourced by topology, we set $c_{\vec{0}}=0$ and take at least two of the $\beta_i$ to be unequal henceforth. For a minimally coupled scalar field, the Casimir stress-energy tensor retains the same anisotropic lattice structure but, as shown in \cref{eq:Tminim2}, acquires additional contributions proportional to $\eta^{2}$, which dominate over the terms proportional to $\eta^{4}$ towards the end of inflation. The physical consequences of the minimally coupled Casimir contribution, together with the role of the isotropic term, will be analyzed in future work.

\color{black}{W}e consider a de Sitter expanding phase as the background solution and treat the topology-induced Casimir stress-energy tensor as a small correction to it. We therefore decompose the Einstein equations into zeroth- and first-order contributions,
\begin{eqnarray}
    &\mathrm{0^{th}}:& \bar{R}_{\mu}{}^{\nu} - \frac{1}{2} \bar{R} \delta_{\mu}{}^{\nu} + 3 H^2 \delta_{\mu}{}^{\nu} =0\,,  \\ \label{eq:first}
     &\mathrm{1^{st}}:&  R^{(1)}_{\mu}{}^{\nu} - \frac{1}{2} R^{(1)} \delta_{\mu}{}^{\nu} = 8 \pi G 
     \langle T^{\mathrm{tp}}{}_{\mu}{}^{\nu}\rangle \,.
\end{eqnarray}
The zeroth-order equations admit the de Sitter solution given in \cref{eq:bkmetric}, while the first-order correction is obtained by linearizing $R_{\mu}{}^{\nu}$ and $R$ around the background de Sitter metric to first order in the small anisotropic perturbations $\delta a_j$ using
\begin{equation}\label{eq:1stmetric}
\mkern-12mu g_{\mu \nu} =
\frac{1}{H^2 \eta^2}
\left[\eta_{\mu\nu} + \operatorname{diag}(0,\delta a_1(\eta),\delta a_2(\eta),\delta a_3(\eta))\right] \,.
\end{equation}
In \cref{eq:1stmetric}, the perturbation of the $00$-component is absorbed by a time reparameterization. This gauge choice prevents us from eliminating any of the spatial perturbations $\delta a_j$, leaving only a residual freedom that rescales their absolute values. Hence only relative differences between the $\delta a_j$ are physical, and local isotropy cannot in general be preserved since the Casimir stress-energy tensor in \cref{eq:Tmunu} sources anisotropic expansion. Defining  
\begin{eqnarray}
    &Z(2) &\equiv \sum_{\vec{r}  \in \mathbb{Z}^3 \setminus \{\vec{0}\} } \frac{1}{\left(\beta_1^2r_1^2+\beta_2^2r_2^2+\beta_3^2 r_3^2\right)^{2}} \,, \\
    &X_j &\equiv  \sum_{\vec{r}  \in \mathbb{Z}^3 \setminus \{\vec{0}\} } \frac{\beta_j^2 r_j^2}{\left(\beta_1^2 r_1^2+\beta_2^2 r_2^2+\beta_3^2 r_3^2\right)^3} \,,
\end{eqnarray}
we integrate \cref{eq:first} and obtain, for each spatial direction,
\begin{equation}\label{eq:solutions}
\delta a_j=\frac{G H^2 \eta^4}{\pi L^4}\left[3 Z(2)-8 X_j\right]+\tilde{d}_j \eta^3+\bar{d}_j\, .
\end{equation}
Here $\tilde{d}_j$ and $\bar{d}_j$ denote the two integration constants associated with the $j$-th scale factor perturbation, which are not all independent due to the $G_{00}$ equation. The resulting metric in \cref{eq:1stmetric}, together with the solutions in \cref{eq:solutions}, represents the first-order correction induced by Casimir backreaction in a toroidal universe with a generic rectangular fundamental domain. With this result in hand, we now quantify how topology-induced quantum effects influence the dynamics of inflation in a representative setting that makes the physical mechanism transparent.

\begin{figure*}[t]
\centering
\includegraphics[width=1\textwidth]{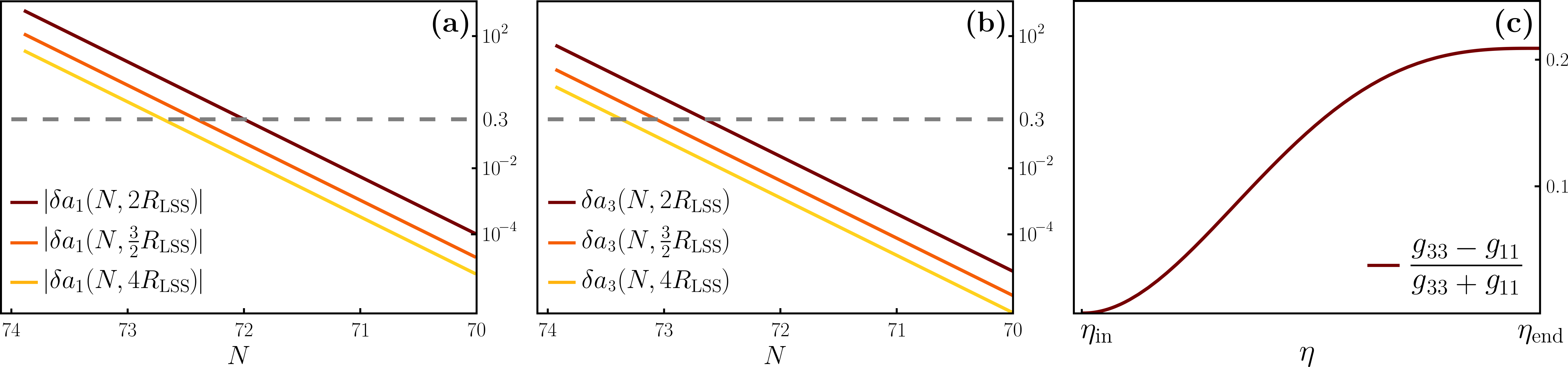}
\caption{\footnotesize{Anisotropy generated purely by quantum Casimir backreaction in a three-torus with rectangular fundamental domain. Panels (a) and (b) show, on a logarithmic scale, the quantities $|\delta a_1(N)|$ and $\delta a_3(N)$ for different values of the fundamental-domain size $L$ as functions of the number of $e$-folds $N$.  dashed lines mark the threshold beyond which the perturbative description is no longer valid. Panel (c) displays the fractional anisotropy between the $11$-component and $33$-component of the first-order-corrected metric, starting from zero at the beginning of inflation, $\eta=\eta_{\mathrm{in}}$, (as imposed by the initial conditions) and growing to a maximum of order $\sim 20 \%$ by the end of inflation. In all panels we take $\beta_1=\beta_2=1$, $\beta_3=2$, $8\pi GH^2=10^{-10}$, and $\eta_{\mathrm{in}}=- e^{N} 10^{14}\,\mathrm{GeV}^{-1} $. In (a) and (b) we evaluate the corrections at $\eta=\eta_{\mathrm{end}}=- 10^{14}\,\mathrm{GeV}^{-1} $. In (c) we show the limiting case $L=2R_{\mathrm{LSS}}=4\cdot10^{42}\, \mathrm{GeV}^{-1}$ and $N=72$, chosen to remain within the perturbative regime.}}
\label{fig:plots}
\end{figure*}

\color{black}{\para{Imprints of topology on early-Universe expansion.}} 
We next fix the initial conditions, specify a particular fundamental domain, and choose representative cosmological parameters. To ensure that any anisotropy is dynamically generated by non-trivial topology rather than imposed in the initial state, we choose initial conditions that coincide with the homogeneous and isotropic de Sitter solution. Accordingly, we impose
\begin{equation}\label{eq:InitialCond}
\delta a_i(\eta_{\mathrm{in}})=0\,; \, \delta a_1'(\eta_{\mathrm{in}}) =\delta a_2'(\eta_{\mathrm{in}}) =\delta a_3'(\eta_{\mathrm{in}}) \, ,
\end{equation}
where $\eta_{\mathrm{in}}$ denotes an initial time. The five initial conditions in \cref{eq:InitialCond}, combined with the constraint imposed by the $00$ component of the first-order Einstein equations, uniquely determine the six integration constants in \cref{eq:solutions}, up to the overall gauge-dependent normalization of the $\delta a_j$. We then consider $\beta_1=\beta_2=1$ and $\beta_3=2$, an elongated fundamental domain for which the 3-direction is twice as long as the 1- and 2-directions. We take $8\pi GH^2=10^{-10}$ and, to quantify the growth of the anisotropic corrections during inflation, we parametrize the solution in terms of the number of $e$-folds $N$. We therefore write the initial time as $\eta_{\mathrm{in}}=e^N\eta_{\mathrm{end}}$. Normalizing the present-day scale factor to unity and assuming temperatures $T_{\mathrm{t}}=10^{-13}\,\mathrm{GeV}$ today and $T_{\mathrm{reh}}=10^{15}\,\mathrm{GeV}$ at reheating, we find that inflation ends at $\eta_{\mathrm{end}}=-10^{14}\mathrm{GeV}^{-1}$,  so that  $\eta_{\mathrm{in}}=-e^{N}10^{14}\mathrm{GeV}^{-1}$. Finally, we choose $L$, the comoving length scale of the fundamental domain, to be greater than or equal to $2R_{\mathrm{LSS}}$, the present diameter of the LSS, thereby satisfying current constraints on the distance to our nearest clone. In this representative setup, \color{black}{we} obtain $\delta a_1(\eta,N)=\delta a_2(\eta,N)$ and a distinct $\delta a_3(\eta,N)$, shown on a logarithmic scale in panels (a) and (b) of \cref{fig:plots} as functions of the number of $e$-folds. As the total number of $e$-folds of inflation increases, the initial volume of the Universe decreases, and so the anisotropic corrections to the stress-energy tensor increase. Physically this reflects that the earlier inflation begins, and hence the smaller the initial size of the Universe, the larger the contribution from the topology-induced stress tensor $\langle T^{\mathrm{tp}}{}_{\mu}{}^{\nu} \rangle$. For sufficiently large $N$, the corrections become comparable to the background geometry, signaling a breakdown of the perturbative description. Enforcing $|\delta a_j(\eta_{\rm end})|\leq0.3$, shown as dashed lines in panels (a) and (b) of \cref{fig:plots}, restricts the allowed number of $e$-folds for fixed $(L,\beta_j,H)$. In particular, for the limiting case $L=2R_{\mathrm{LSS}}$, perturbativity requires $N\lesssim 72$.

To illustrate the impact of Casimir backreaction during inflation, panel (c) of \cref{fig:plots} shows the fractional anisotropy between the first-order-corrected spatial metric components $g_{11}$ and $g_{33}$. In an exactly isotropic de Sitter spacetime, this quantity vanishes, so any deviation from zero directly measures the difference between the effective expansion rates along the $1$- and $3$-directions. 
Starting from zero by construction, the anisotropy grows most rapidly in the very early Universe, near $\eta \gtrsim \eta_{\mathrm{in}}$, when the sourcing term is most dynamically relevant, such that by the end of inflation the expansion rates along the $1$- and $3$-directions differ by $\sim 20\%$.
Towards the end of inflation the curve flattens, showing the usual smoothing effect of inflation: the anisotropic Casimir source is diluted and becomes progressively less important relative to the isotropic expansion sourced by the cosmological constant. Such smoothing, however, does not necessarily erase all traces of the Casimir-driven anisotropic phase.  Modes that leave the horizon during this phase are expected to retain direction-dependent corrections in the primordial power spectrum, even after the subsequent expansion becomes increasingly isotropic. The positive value of the fractional anisotropy in panel (c) of \cref{fig:plots} shows that the $1$- and $2$- directions expand slower than the $3$-direction. Consequently, since the $3$-direction is initially longer than the other two, the induced anisotropic expansion enhances this initial hierarchy.
Panel (c) of \cref{fig:plots} therefore demonstrates that, even within the perturbative regime, Casimir backreaction provides a concrete mechanism by which non-trivial topology induces departures from a perfectly isotropic background expansion in the early Universe.

\para{Conclusions.} In this work we presented a concrete semiclassical-gravity setting in which a purely quantum effect, the Casimir stress-energy tensor, produces signatures that can be connected to cosmological observables. We have shown that, independent of the specific field content, scheme-independent Casimir contributions naturally arise in universes with non-trivial topology. Such contributions encode locally the symmetries broken globally by the underlying topology. By breaking de Sitter invariance, non-trivial topology removes the usual symmetry justification for setting the zero-mode contribution to zero, allowing a potentially physical, topology-dependent isotropic contribution to the effective equation of state. Furthermore, in the representative case of a flat FLRW background with three-torus topology and a rectangular fundamental domain, we have shown that Casimir backreaction induces anisotropic early-Universe expansion, thereby demonstrating how a universe with non-trivial topology can differ from its simply connected counterpart already at the level of the background expansion. In more generic topologies, reflecting further reductions in global symmetry, Casimir effects are expected to induce both anisotropic and inhomogeneous expansion.

The presented mechanism has no classical analogue: Casimir contributions arise from the boundary conditions imposed by global topology and are therefore an unavoidable consequence of quantum fields in a multiply connected universe. Conversely, if the spatial topology of the Universe were established to be non-trivial, evaluating Casimir backreaction, under the requirement of an approximately homogeneous and isotropic CMB, would place non-trivial consistency conditions on the duration of inflation.
In upcoming work, we will extend the present analysis by including the quantum creation and evolution of perturbations, determining whether modes exiting the horizon during the topology-driven anisotropic phase retain a detectable directional dependence. The background-level anisotropy found here provides a concrete semiclassical link between cosmic topology and cosmological observables: it lays the foundation for quantitative predictions for CMB observables, putting on a firmer footing the long-standing studies of the effects of topology on the CMB \cite{Starobinsky:1993yx,deOliveira-Costa:1995vur,Cornish:1997hz,Cornish:1997rp,Levin:1997tu,Weeks:1998qr,starobinsky1993newrestrictionsspatialtopology,PhysRevLett.71.20,Cornish:1997ab} and opening a window on additional quantum effects in the early Universe.

\begin{acknowledgments}
We thank B.\ Fornal for helpful comments that improved the manuscript. 
A.N.\ is supported by the Richard S.\ Morrison Fellowship.  
G.D.S. and K.H. are supported by DOE award DESC0009946.
Y.A. acknowledges support by the Spanish Research Agency (Agencia Estatal de Investigaci\'on)'s grant RYC2020-030193-I/AEI/10.13039/501100011033, by the European Social Fund (Fondo Social Europeo) through the  Ram\'{o}n y Cajal program within the State Plan for Scientific and Technical Research and Innovation (Plan Estatal de Investigaci\'on Cient\'ifica y T\'ecnica y de Innovaci\'on) 2017-2020, by the Spanish Research Agency through the grant IFT Centro de Excelencia Severo Ochoa No CEX2020-001007-S funded by MCIN/AEI/10.13039/501100011033, by the Spanish National Research Council (CSIC) through the Talent Attraction grant 20225AT025, and by the Spanish Research Agency's Consolidaci\'on Investigadora
2024 grant CNS2024-154430.
J.C.D. is supported by the Spanish Research Agency (Agencia Estatal de Investigaci\'on), the Ministerio de Ciencia, Inovaci\'on y Universidades, and the European Social Funds through grant JDC2023-052152-I, as part of the Juan de la Cierva program.
M.M.B. acknowledges support by the Spanish Ministry of Science, Innovation and Universities under the FPU predoctoral grant FPU22/02306.
T.S.P is supported by Funda\c{c}\~ao Arauc\'aria (NAPI Fen\^omenos Extremos do Universo, grant 347/2024 PD\&I).
G.A. is supported by the Spanish Research Agency's Consolidaci\'on Investigadora
2024 grant CNS2024-154430.
F.C.G\ was supported by the Presidential Society of STEM Postdoctoral Fellowship at Case Western Reserve University and by Ministerio de Ciencia, Innovaci\'on y Universidades, Spain, through a Beatriz Galindo Junior grant BG23/00061.
D.P.M. acknowledges support by the Bulgarian National Science Fund program ``VIHREN--2024" project No. KP--06--DV/9/17.12.2024.
J.J.O.G acknowledges support by the Spanish Ministry of Science, Innovation and Universities under the FPU predoctoral grant FPU24/02041.
A.T. is supported by the European Union's Horizon Europe research and innovation programme under the Marie Sk\l odowska-Curie grant agreement No. 101126636.
J.C.D. is and A.T. was supported by CSIC through grant No. 20225AT025.
G.D.S., C.J.C., A.K. and D.P.M.\ acknowledge partial support from NASA ATP grant RES240737 and from NASA ADAP grant 24-ADAP24-0018;  A.H.J.\ from the Royal Society and STFC in the UK; and A.S. from DOE grant DESC0009946. We are thankful to the Istituto Nazionale di Fisica Nucleare in Italy.
\end{acknowledgments}

\bibliographystyle{apsrev4-1}
\bibliography{main}

\end{document}